\documentclass[a4paper]{jpconf}

\usepackage{graphicx,iopams}

\begin{document}

  \title{Stochastic Green's function approach to disordered systems}
  \author{A. Alvermann and H. Fehske}
  \address{Institut f\"ur Physik, Ernst-Moritz-Arndt-Universit\"at, 17487 Greifswald, Germany}

  \date{\today}

  \begin{abstract}
    Based on distributions of local Green's functions
    we present a stochastic approach to disordered systems.
    Specifically we address Anderson localisation and cluster effects
    in binary alloys.
    Taking Anderson localisation of Holstein polarons as an example
    we discuss how this stochastic approach can be used for the investigation
    of interacting disordered systems.
  \end{abstract}

  \section{Introduction}

  In many cases the influence of crystal imperfections present in any real
  material can be neglected in favour of the interactions dominating
  e.g. electron transport or optical properties.
  In some cases however, and especially at low temperatures,
  interesting physical effects arise from the specific
  physical processes induced by scattering on crystal defects.  
  Examples range from the quantum Hall effect --
  an intricate problem far beyond the scope of our study
  -- to electrical transport in polaronic systems.
  One prominent example is Anderson's prediction that in a
   disordered material itinerant (extended) states can be
  turned into localised states which do not carry any current~\cite{anderson58}.
  This transition from extended to localised states 
  is a result of quantum interference arising from
  elastic scattering on impurities in a crystal.
  It leads, roughly speaking, to spatial confinement of an electron:
  The electron wave-function decays exponentially with distance.
  
  This peculiar, essentially quantum mechanical behaviour has motivated 
  lots of research (see e.g.~\cite{kramer93} for a review).
  As yet, however, many important questions in the field of localisation
  physics have not been finally settled, and 
  localisation in interacting systems is still far from being
  understood.
  Progress on this topic requires new techniques which allow for
  a comprehensive -- necessarily approximate, but reliable --
  description of disorder and interaction on a microscopic scale.
  One possible approach shall be discussed here.

  The focus of our study is on substitutionally
  disordered three-dimensional materials like doped semiconductors or alloys.
  We will not be concerned with amorphous materials like glasses
  which do not possess a crystal lattice (but see
  e.g.~\cite{logan84,logan87}).
  Then disorder primarily manifests through site-dependent local potentials
  $\epsilon_i$.
  The motion of an electron in such a disordered crystal is described
  by the tight-binding Hamiltonian
  \begin{equation}\label{hamilton}
    H = \sum_i \epsilon_i c^\dagger_i c_i -t \sum_{\langle i, j \rangle}
    c^\dagger_i c_j \;.
  \end{equation}
  In a perfect crystal, i.e. $\epsilon_i=0$, the electron hopping $t$
  between nearest--neighbour sites $\langle i, j \rangle$ 
  gives rise to a band of width $W_0$,
  e.g. on a cubic lattice $W_0 = 12 t$, or
   $W_0 = 4 t \sqrt{K}$ on a Bethe lattice with connectivity $K$.
  In what follows, we fix the bandwidth $W_0=1$, 
  and measure energies in units of
  $W_0$.

  To model the stochastic character of disorder
  the $\epsilon_i$ are considered as random variables with a given
  distribution.
  We demand that they are independently identically
  distributed with common distribution $P(\epsilon_i)$.
  For such a stochastic Hamiltonian 
  quantities like the (retarded) Green's function
  $G_{ij}(\omega)= \lim_{\eta\to0^+}\langle 0 |c_i \left[\omega+\mathrm{i}\eta-H
  \right]^{-1} c^\dagger_j |0 \rangle$ are random variables in their own right.
  It is thus reasonable to ask for their distributions.
  This is especially true for the local density of states (LDOS)
  \begin{equation}\label{def:LDOS}
    \rho_i(\omega)=- \mathrm{Im}\; G_{ii}(\omega)/\pi \;.
  \end{equation}
  Its distribution $P[\rho_i(\omega)]$ 
  gives the statistics of the LDOS 
  in a disordered system,
  where translational symmetry is broken and $\rho_i(\omega)$ varies
  with $i$.
  Note that $P[\rho_i(\omega)]$ does not depend on the site index $i$:
  Translational symmetry is restored on the level of distributions.
  It is important to realise that the site-dependence of the LDOS
  constitutes an eminent aspect of a disordered system.
  For an extended state, for example, $\rho_i(\omega)$ has a finite value
  on most lattice sites.
  For a localised state, in contrast, $\rho_i(\omega)$ has
  exponentially small values on lattice sites outside a finite region,
  corresponding to the decay of the wave-function. 
  $P[\rho_i(\omega)]$ will thus be different for extended and
  localised states.
  Obviously the disorder averaged density of states (DOS)
  \begin{equation}
    \rho_\mathrm{ave}(\omega) = \langle \rho_i(\omega) \rangle =
    \int\limits_0^\infty
    \rho_i \; P[\rho_i(\omega)] \; d\rho_i \;,
  \end{equation}
  merely counting the number of states at a given energy $\omega$,
  cannot account for this signature.
  Only for the ordered case $P(\epsilon_i)=\delta(\epsilon_i)$, when
  translational symmetry implies that $\rho_i$ does not depend on $i$,
  the distribution is entirely determined by the DOS,
  $P[\rho_i(\omega)] = \delta[\rho_i-\rho_\mathrm{ave}(\omega)]$.
  As we will later see a similar
  characterisation does not even hold approximatively in a disordered system. 
  The full distribution can be extremely broad for 
  states in which the electron strongly scatters at impurities,
  catching the resulting deviations in $\rho_i$.

  \section{Local distribution approach}
  Our goal is to obtain a calculational scheme for the distribution of
  the LDOS for the models given by Eq.~(\ref{hamilton}).
  This scheme must account for the correlations that
  make up Anderson localisation.
  At the same time it should be extendable to incorporate interactions. 
  Naturally this requires to employ approximations.
  It is important to guarantee that within these approximations 
  the mutual influence of interaction and disorder at the microscopic scale
  is still represented (which rules out e.g. the coherent potential
  approximation (CPA), see below).
  A definite candidate to meet these demands
  is what we call the local distribution (LD) approach.
  It provides a self-consistent scheme for the distribution
  $P[G_{ii}(\omega)]$ of the local Green's function $G_{ii}(\omega)$.
  We first describe this approach and its application to disordered
  systems, and later will combine it with a treatment of interaction.
  
  \subsection{LD approach on a Bethe lattice}

\begin{figure}
    \begin{minipage}{0.4\linewidth}
      \hspace*{1cm}
      \includegraphics[width=0.5\linewidth]{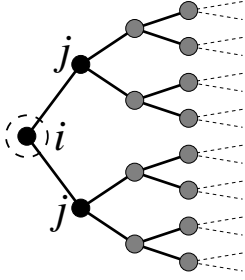}
    \end{minipage}
    \hfill
    \begin{minipage}{0.5\linewidth}\vspace*{1.5cm}
    \caption{Sketch of the (half-infinite) Bethe lattice with
      coordination number $K=2$.
      When site $i$ is removed from the lattice the sites $j$ are in 
      the same geometrical situation as site $i$ before, and no path
      connects them.}
    \label{FigBethe}
    \end{minipage}
  \end{figure}

  It is straightforward to construct the LD approach following the work of 
  Abou-Chacra, Anderson  and Thouless~\cite{Abou73}.
  We briefly repeat their derivation which is carried out on a
  Bethe lattice (see Fig.~\ref{FigBethe}).
  The basic observation is the decomposition of the local Green's
  function $G_{ii}(\omega)$,
  \begin{equation}\label{GBethe}
    G_{ii}(\omega) = \left[\omega - \epsilon_i - t^2 \sum_{j=1}^K
      G^{(i)}_{jj}(\omega) \right]^{-1} \;.
  \end{equation}
  The sum runs over the $K$ neighbours $j$ of $i$,
  and the Green's functions $G^{(i)}_{jj}(\omega)$ have to be
  calculated for the lattice after site $i$ is removed.
  For the ordered case we have 
  $G^{(i)}_{jj}=G_{ii}(\omega)$ (cf. Fig~\ref{FigBethe}),
  and Eq.~(\ref{GBethe}) is a quadratic equation for the
  Green's function. Especially we can read off the semi-circular DOS 
  $\rho(\omega) = 4/(\pi W_0^2) \sqrt{W_0^2-4\omega^2} $ of the Bethe
  lattice~\cite{economou83}.
  
  Now the stochastic viewpoint comes into play. For a disordered system, 
  like $\epsilon_i$ all Green's functions are stochastic quantities,
  i.e. random variables, and Eq.~(\ref{GBethe}) determines the random variable 
  $G_{ii}(\omega)$ in terms of $K+1$ others, namely $\epsilon_i$ and
  $G^{(i)}_{jj}(\omega)$.
  On the Bethe lattice the Green's functions $G^{(i)}_{jj}(\omega)$
  correspond to the same geometrical situation as $G_{ii}(\omega)$,
  hence they are identically distributed.
  Furthermore no path connects the sites $j=1,\dots,K$ if $i$ is
  removed. This implies that the $G^{(i)}_{jj}(\omega)$ are
  independently distributed.
  Therefore Eq.~(\ref{GBethe}) can be regarded as a self-consistency
  equation which determines the distribution of one random variable,
  the local Green's function $G_{ii}(\omega)$, or equivalently
  $G^{(i)}_{jj}(\omega)$.
  The distribution $P(\epsilon_i)$ enters this equation as a parameter.
  
  Such a stochastic self-consistency equation can be numerically
  solved by a Monte-Carlo procedure (Gibbs-sampling).
  To this end we represent the random variable $G_{ii}(\omega)$
  through a sample of $N$ entries (typically $N=10^4 \dots 10^7$).
  Each entry of the sample is repeatedly updated by a new value from
  Eq.~(\ref{GBethe}), with $K$ values for $G^{(i)}_{jj}(\omega)$
  randomly drawn from the sample and a randomly chosen $\epsilon_i$.
  The distribution of $G_{ii}(\omega)$ is finally constructed like a histogram
  by counting the number of entries in the sample with a specific value.

\subsection{LD approach on arbitrary lattices}

To put the LD approach in a more general context
we will show how to construct it as an approximative scheme 
on others than the Bethe lattice, thereby making contact to the
CPA (concerning CPA see e.g.~\cite{elliott74}).
On an arbitrary lattice the Green's function can still be decomposed as
\begin{equation}\label{Giidecomp}
  G_{ii}(\omega) = \left[ \omega - \epsilon_i - t^2 \sum_{l,j=1}^K
    G^{(i)}_{lj}(\omega) \right]^{-1} \;.
\end{equation} 
In contrast to the Bethe lattice the $G^{(i)}_{lj}(\omega)$ which
appear in this decomposition do not correspond to the same geometrical
situation as $G_{ii}(\omega)$.
If we iterated Eq.~(\ref{Giidecomp}) we would thus end up with an
infinite hierarchy of equations containing many different Green's
functions.
As an immediate consequence the LD approach, relying on local Green's
functions, will provide only an approximative scheme on general
lattices. 

A simple approximation is suggested by introducing 
\begin{equation}
  H^{(i)}_j(\omega) = t^2 \sum_{l=1}^K G^{(i)}_{lj}(\omega)
\end{equation}
which is the part of the $K$ hopping contributions in Eq.~(\ref{Giidecomp})
where the electron leaves site $i$ through $j$.
Specifically we assume that, for each $j$, $H^{(i)}_j(\omega)$ is determined
by the local Green's function $G_{jj}(\omega)$ alone. 
Moreover, the $K$ Green's functions $G_{jj}(\omega)$ are
taken to be independently distributed. 

To proceed further we write the local Green's function 
with a self-energy like in the CPA 
\begin{equation}\label{Giisig}
  G_{ii}(\omega)=G^0(\omega-\Sigma_i(\omega)) \;,
\end{equation}
where $G^0(\omega)$ is the `bare' propagator for the ordered system.
In contrast to the CPA the self-energy $\Sigma_i(\omega)$ is now
site-dependent.
Thinking in terms of an effective medium in which site $i$ is embedded
this medium is characterised by the
distribution of self-energies $\Sigma_i(\omega)$.
The electron at site $i$ 
`sees' the effective medium through the $K$ Green's functions
$G_{jj}(\omega)$.
Being part of the effective medium
the hybridisation $H^{(i)}_j(\omega)$ associated with $G_{jj}(\omega)$
must fulfil 
$G_{jj}(\omega) = \left[\omega-\Sigma_j - K H^{(i)}_j(\omega)\right]^{-1}$.
Inverting this equation yields 
\begin{equation}\label{Hjj}
  H^{(i)}_j(\omega) = \frac{1}{K}\left(\omega-\Sigma_j-\left[G_{jj}(\omega)\right]^{-1}\right) \;.
\end{equation}

The Green's function $G_{ii}(\omega)$ is then given by 
$K$ other Green's function $G_{jj}(\omega)$ of the same type:
\begin{equation}\label{GiifromGjj}
  G_{ii}(\omega) = \left[\omega-\epsilon_i-\sum_{j=1}^K H^{(i)}_j(\omega)
  \right]^{-1} =
  \left[ \frac{1}{K} \sum_{j=1}^K
    \left[G_{jj}(\omega)\right]^{-1} - \epsilon_i + 
    \frac{1}{K} \sum_{j=1}^K \Sigma_j(\omega) \right]^{-1} \;.
\end{equation}
Clearly, if all $\epsilon_i=0$, this equation reduces to $\Sigma_i=0$,
that is $G_{ii}(\omega) = G^0_{ii}(\omega)$.

At last, we have a complete set of equations, which with the same
Monte-Carlo-scheme as before can be solved to determine the distribution of
$G_{ii}(\omega)$.
Note that the approximations we made are guaranteed to respect causality,
in the sense that $\mathrm{Im}\, G_{ii}(\omega),\, \mathrm{Im}\,
\Sigma_{i}(\omega) < 0$.
Eqs.~(\ref{Giisig}) and ~(\ref{GiifromGjj}) are immediately seen to
preserve this important property.
For Eq.~(\ref{Hjj}), where this property is not obvious,
we can rely on a result from the analyticity proof of the
CPA~\cite{MHart73}, which states that $\mathrm{Im}\left\{ \Sigma +
\left[G^0(\omega-\Sigma)\right]^{-1} \right\} > 0 $ for $\mathrm{Im}\,\Sigma <0$.

We like to point out that the construction done here is slightly
ambiguous.
A different approximation in Eq.~(\ref{Hjj}) 
would still give a self-consistent scheme for distributions of local
Green's functions $G_{ii}(\omega)$.
Anyhow, our scheme incorporates the CPA and reproduces the exact
scheme for the Bethe lattice in a natural way.
The CPA results from Eq.~(\ref{GiifromGjj})
for $K=\infty$ by the central limit theorem.
Then, the sums over Green's functions and self-energies are replaced
by averages, yielding the CPA equation for the averaged local Green's
function
\begin{equation}
 G^\mathrm{CPA}(\omega) = \left\langle
   \frac{1}{\left[G^\mathrm{CPA}(\omega)\right]^{-1} -
     (\epsilon_i -\Sigma(\omega))} \right\rangle \;.
\end{equation}
In principle, $K$ is in our scheme no free parameter but the
number of neighbours to a lattice site.
If we increase $K$ the bare propagator $G^0(\omega)$ changes,
and therefore we shall scale $t$ as $t \propto 1/\sqrt{K}$ like in the limit of high
dimension~\cite{MV89}.
Then, for $K\to\infty$, our scheme still reduces to the CPA, which is known
to become exact for lattices with infinite connectivity~\cite{VV92}.
This implies that the approximations we made are good for high-dimensional 
lattices. Actually they turn out to be good for $d \ge 3$.
It is commonly known that
CPA is the best single-site approximation for disordered systems.
The local Green's function is obtained on average from an `averaged'
effective medium.
The LD approach solves the local problem exactly,
while the hybridisation is calculated from a `fluctuating' effective medium.
So to say, the LD approach is the best single-site approximation if
fluctuations are taken into account.

On the Bethe lattice the Green's function fulfils
$G^0_{ii}(\omega) = \left[\omega - t^2 K G^0_{ii}(\omega)\right]^{-1}$
(this is just Eq.~(\ref{GBethe}) for $\epsilon_i=0$),
whereby Eq.~(\ref{Giisig}) reads 
\begin{equation}
  \Sigma_i(\omega) - \left[ G_{ii}(\omega) \right]^{-1} 
  = \omega-t^2 K G_{ii}(\omega)
\end{equation}
Then Eq.~(\ref{GiifromGjj}) reduces to Eq.~(\ref{GBethe}) for the
Bethe lattice. 
The key point is of course that, owing to the specific Bethe lattice geometry,
the non-diagonal contributions $G^{(i)}_{jl}$ are zero,
and the $H^{(i)}_j$ are indeed independently distributed.
Then our approximation scheme becomes exact: On the Bethe lattice the
best single-site approximation, with fluctuations, is an exact theory.
Note that for the Bethe lattice $K$ is a free parameter:
The `free' DOS in absence of disorder is, for fixed $W_0$, independent of $K$,
but choosing $K$ fixes the Bethe lattice used.

As the transfer of the LD approach to other lattices shows,
going to the Bethe lattice basically changes the `free' DOS
and introduces approximations to cubic lattices by
neglecting correlations. The essential properties of disordered
systems, e.g. localisation, will however be well described.
For the time being we work on a $K=2$ Bethe lattice.

\section{Disordered crystals}

Below we will discuss two models,
the Anderson model with a `continuous' probability
distribution,
and the binary alloy model with a `discrete' one.
While the former is dominated by the Anderson transition from extended to
localised states, the latter is dominated by 
multiple electron scattering on clusters of atoms, visible e.g. in the
fragmentation of the DOS (see below).

\subsection{Anderson model}

In the Anderson model, with disorder strength $\gamma \ge 0$, the $\epsilon_i$ have a box distribution
\begin{equation}\label{EpsAnd}
  P(\epsilon_i) = (1/\gamma)\, \Theta(\gamma/2-|\epsilon_i|) \;.
\end{equation}

\begin{figure}
  \begin{minipage}{0.65\linewidth}
    \hspace*{1cm}
    \includegraphics[width=0.7\linewidth]{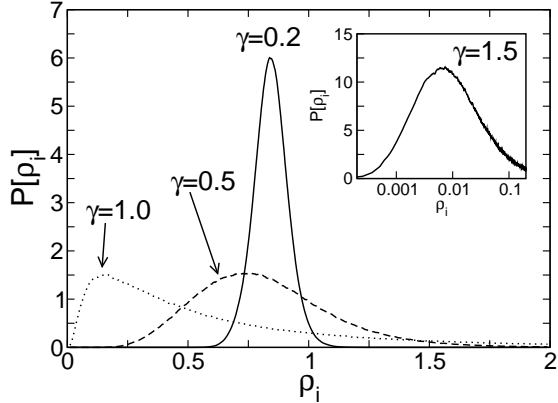} 
  \end{minipage}
  \hfill
  \begin{minipage}{0.34\linewidth}\vspace*{1.5cm}
    \caption{Change of the probability distribution
      $P[\rho_i(\omega)]$ of the LDOS $\rho_i(\omega)$ for the
      Anderson model, in the band
      centre $\omega=0$, with increasing $\gamma$. 
      Note that the inset has a logarithmic $\rho_i$-axes.}
    \label{AndDist}
  \end{minipage}
\end{figure}

\begin{figure}
  \begin{minipage}{0.65\linewidth}
    \hspace*{1cm}
    \includegraphics[width=0.7\linewidth]{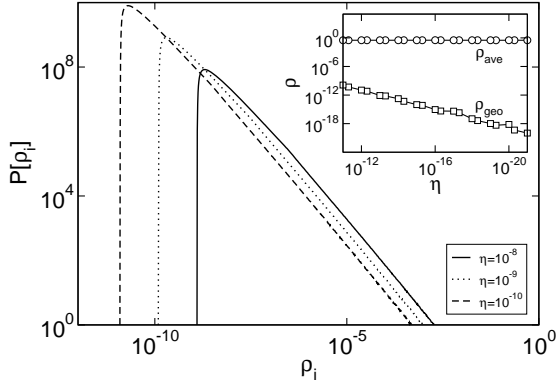}
  \end{minipage}
  \hfill
  \begin{minipage}{0.34\linewidth}
    \caption{The figure displays, for localised states 
      ($\gamma=1.5$, $\omega=0.9$), the change of $P[\rho_i]$
      with decreasing $\eta$.
      The inset shows that $\rho_\mathrm{ave}$ is constant for
      $\eta \to 0$, but a typical moment,
      the geometrically averaged DOS
      $\rho_\mathrm{geo} = \exp \langle \ln \rho_i \rangle$,
      goes to zero.  
    For extended states we get distributions like in
    Fig.~\ref{AndDist}, independent of (a small) $\eta$.}
  \label{AndEta}
  \end{minipage}
\end{figure}

\begin{figure}
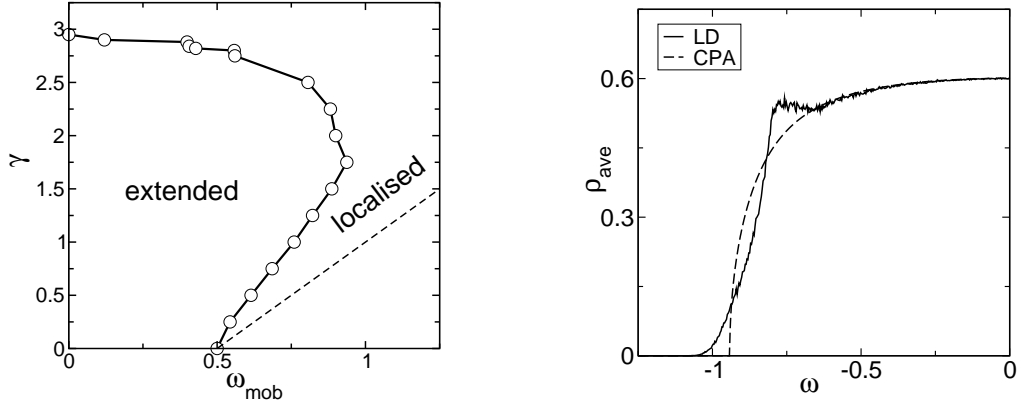

  \hfill
  \hspace*{1cm}
  \begin{minipage}{0.4\linewidth}
    \includegraphics[width=0.9\linewidth]{andphas} 
  \end{minipage}
  \hspace*{1cm}
  \begin{minipage}{0.4\linewidth}
    \includegraphics[width=0.9\linewidth]{anddos} 
  \end{minipage}
  \hspace*{1cm}
  \caption{
      Left panel: Mobility edge
      trajectory, i.e. the phase diagram,
      of the Anderson model, obtained in the limit $\eta\to 0$. 
      The dashed lines indicate the band edges $\pm (W/2 +\gamma/2)$.
      Right panel: DOS
      for the Anderson model at $\gamma=1.5$ for $\omega \le 0$,
      calculated with CPA and LD approach. 
      Both the DOS and the phase diagram are symmetric under
      $\omega\to -\omega$.}
    \label{AndPhasDOS}
\end{figure}

Fig.~\ref{AndDist} shows the change of the distribution
$P[\rho_i(\omega)]$ of the LDOS
with increasing disorder, as obtained from our Monte-Carlo scheme.
We find the behaviour indicated in
the introduction. For small disorder the distribution is rather
Gaussian, but becomes strongly asymmetric and extremely broad as
$\gamma$ increases. The (averaged) DOS $\rho_\mathrm{ave}(\omega)$ is
not indicative of these distributions.
If $\gamma$ exceeds a critical value $\gamma_c(\omega)$ 
[$\gamma_c(\omega=0) \simeq 3.0$] the distribution becomes singular while
$\rho_\mathrm{ave}(\omega)$ is still finite.
At this point the transition from extended to localised states takes
place.
To determine $\gamma_c(\omega)$ numerically we have to 
study the scaling of appropriate quantities, like for any phase
transition.
Here we employ the different behaviour of $P[\rho_i(\omega)]$ when the
imaginary part $\eta$ in the energy argument of
$G_{ii}(\omega+\mathrm{i}\eta)$ goes to zero:
For extended (localised) states the distribution is stable (instable)
for $\eta\to 0$ (see Fig.~\ref{AndEta}).
This behaviour is again not caught by 
$\rho_\mathrm{ave}(\omega)$, which is finite for $\eta\to0$
for both extended and localised states.
But suitable moments of the distribution,
like the geometric average, drop to zero just for localised states
(see inset in Fig.~\ref{AndEta}). 
With this criterion which numerically tests whether the distribution
is singular or not we can very precisely distinguish extended and
localised states~\cite{A03}.
Fig.~\ref{AndPhasDOS} shows the phase diagram of the
Anderson model on the Bethe lattice obtained by means of the limit
$\eta\to 0$.
It displays two important features of localisation.
First, the existence of a critical disorder $\gamma_c(\omega=0)$
above which all states are localised due to the strong impurity scattering.
Second, before complete localisation occurs,
states towards the band centre are 
extended while states towards the band edges are localised,
these states being separated by the so-called mobility edges.
Furthermore the mobility edge trajectory reveals
that for small disorder the electron tunnels between impurities,
giving rise to extended states outside the band $[-W_0/2,W_0/2]$ of the
ordered system.

Let us finally contrast the LD result for $\rho_\mathrm{ave}(\omega)$
(Fig.~\ref{AndPhasDOS}) with the CPA.
The CPA reproduces the DOS very good inside
the band -- only here $P[\rho_i(\omega)]$ is a Gaussian -- but fails
closer to the band edges where it misses the (Lifshitz) tails in the DOS.
These tails result from states at the band edges which come along
with repeated scattering of the electron on clusters of impurities
with a large (or small) potential $\epsilon_i$.
Since multi-scattering is not accounted for in the CPA these states
cannot be described.

\subsection{Binary alloy}
  
  Multi-scattering becomes rather important for `discrete' distributions like
  \begin{equation}
    P(\epsilon_i) = c_A \delta(\epsilon_i+\Delta/2) +
    (1-c_A) \delta(\epsilon_i-\Delta/2)\,,
  \end{equation}
  describing a two-component binary alloy made of `A-atoms' with potential
$-\Delta/2$ and `B-atoms' with potential $+\Delta/2$.
Thus we have two parameters, the concentration $c_A$ of the A-species
and the energy separation $\Delta$.

\begin{figure}
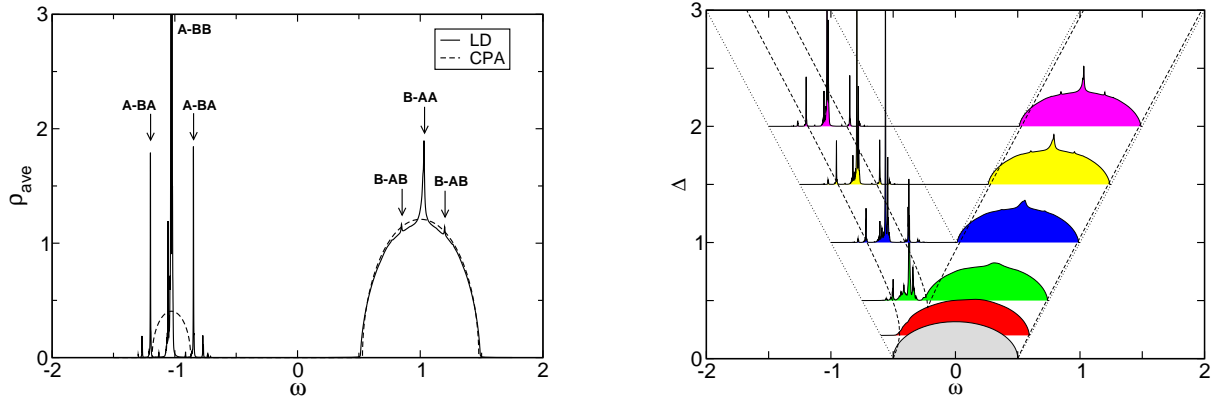

  \includegraphics[width=0.45\linewidth]{BAdos} 
  \hfill
  \includegraphics[width=0.45\linewidth]{BAphas}
  \caption{Left panel: DOS for the binary alloy model at $\Delta=2.0$,
    $c_A=0.1$, contrasting the CPA and LD results.
    Arrows mark different peaks in the DOS which arise from the
    clusters indicated (e.g. \textsf{A-BB} indicates the contribution
    from an A-atom which is surrounded by two B-atoms.)
    We used in the figure an arbitrary $\eta=10^{-3}$ to broaden the peaks.
    Right panel: Kind of a phase diagram for the binary alloy model at
    $c_A=0.1$, showing the DOS for various $\Delta$.
  The dashed curves show the CPA band edges, and the dotted lines mark
  $\omega=\pm \Delta/2 \pm W_0 /2$
  (figures taken from~\cite{AF05})}.
  \label{BADOS}
\end{figure}

Exemplarily we consider the case $c_A=0.1$
when the A-atoms form the minority species in a bulk of B-atoms,
  and $\Delta=2.0$ when the A-atoms
  are energetically well-separated from the band of the B-bulk.
  Fig.~\ref{BADOS} shows the corresponding DOS, contrasting the LD
  and CPA results.
  The B-band is rather smooth and 
  almost reproduced by the CPA, which nevertheless misses important features. 
  In the energy range of the A-atoms we do not find a band but a
  strongly fragmented set of peaks.
  Each of these peaks can be attributed to a specific cluster of A-atoms, as
  indicated in the figure.
  Due to the low concentration $c_A$ of A-atoms, and the large energy
  separation $\Delta$, states at these clusters do not hybridise and
  form a continuous band, but are essentially confined to the clusters
  and contribute discrete peaks to the DOS.
  The CPA is by construction unable to identify those cluster states
  and therefore entirely misses band fragmentation.
  With increasing $\Delta$ these signatures become more prominent,
  while CPA suggests that the bands, being well
  separated, less influence each other (see right panel in Fig.~\ref{BADOS}).
  The CPA is of course able to depict gross features as band
  splitting, but replaces the complex spectrum by semi-circular
  bands with weight $c_A$ and $1-c_A$.
  Then, e.g., the minority A-band has spectral weight outside the CPA
  band: Replacing this band by a semi-circular one while conserving
  the overall spectral weight $c_A$ results in a clearly reduced
  CPA bandwidth.
  
  We think that the binary alloy is a particular instructive example
  how, using the distribution of $G_{ii}(\omega)$, one manages to
  account for multi-scattering events which are absent in a CPA description
  relying on averaged values. 
  With a proper treatment of multi-scattering like in the LD
  approach both quantum interference leading to Anderson
  localisation and formation of cluster states is correctly described.

  \section{Electron-Phonon Coupling}

  Anderson localisation arises from elastic electron-impurity
  scattering.
  Coherence which is maintained during elastic scattering is the
  important precondition for localisation.
  To understand which role localisation plays in a real solid
  we must therefore understand how it is affected by inelastic
  scattering, which naturally arises from the interaction of the
  electron with lattice vibrations, i.e. phonons.
  A model to study the competition between localisation --related to coherence--
  and electron-phonon (EP) interaction --related to incoherence-- is provided by the
  Anderson-Holstein Hamiltonian 
  \begin{equation}
    H = \sum\limits_{i} \epsilon_i c^\dagger_i c_i
      - t
      \sum\limits_{\langle i, j\rangle}  c_{i}^\dagger c_{j}^{} 
      - \sqrt{\varepsilon_p\omega_0} 
      \sum\limits_{i} (b_i^\dagger + b_i^{} ) c^\dagger_i c_i
      + \omega_0 \sum\limits_i b_i^\dagger b_i^{}~,   
   \end{equation} 
  which adds a Holstein-type of interaction to the Anderson model,
  the $\epsilon_i$ being distributed according to Eq.~(\ref{EpsAnd}).
  Here $b_i^{(\dagger)}$ denote bosonic operators describing
  dispersionless optical phonons with frequency $\omega_0$ which are
  locally coupled to the electron density $c_i^\dagger c_i^{}$.
  We introduce the dimensionless phonon frequency $\tilde{\omega}_0 =
  \omega_0 / W_0$, and a coupling constant $\tilde{\lambda}=2
  \varepsilon_p/ W_0$.
  Moreover we consider one electron at $T=0$.
  
  Besides introducing incoherent motion, 
  EP coupling in this model can lead to the formation of a
  new quasiparticle.
  Sufficiently strong EP coupling binds the electron to 
  the lattice deformation at the electron's site,
  forming a new compound entity, the polaron (see e.g.~\cite{Fi75}).
  A polaron is strongly mass-enhanced and therefore very much affected
  by disorder.
  However, a polaron is not just a heavy electron.
  Due to inelastic scattering and retardation of the EP interaction in most
  cases the internal structure of the polaron will play a crucial 
  role, leading to different localisation properties.
  Moreover if inelastic scattering dominates, leading to incoherent
  motion of the polaron, localisation will be suppressed. 
  Then a polaron can be less easy to localise than the free
  electron, although its mass is still increased.
      
  \subsection{Extending the LD approach to interacting systems}

  Like in the standard Green's functions formalism 
  interaction is incorporated in the LD approach by an interaction self-energy
  $\Sigma^I_{ij}(\omega)$~\cite{GJ80}. 
  The `disorder' self-energy $\Sigma_i$, as
  introduced in Eq.~(\ref{Giisig}), is local.
  The interaction self-energy will be approximated to
  be local as well.
  The Green's function $G_{ii}(\omega)$ is then obtained as
  \begin{equation}
    G_{ii}(\omega)\! =\! \left[ \omega-\epsilon_i- \Sigma^I_i(\omega)-\sum_{j=1}^K H^{(i)}_j(\omega) \right]^{-1}\! =\!
  \left[ \frac{1}{K} \sum_{j=1}^K
    \left[G_{jj}(\omega)\right]^{-1} - \epsilon_i - \Sigma^I_i(\omega)
    + \frac{1}{K} \sum_{j=1}^K \Sigma_j(\omega)  \right]^{-1}\!\!\!,
  \end{equation}
  replacing Eq.~(\ref{GiifromGjj}).
  Like for the disorder part one should demand~\cite{dobro98}
  that the interaction self-energy
  is obtained as the `best' local self-energy
  which amounts to use the dynamical mean field theory (DMFT) (for a
  review see~\cite{georges96}).
  Within DMFT the interaction self-energy is a functional
  $\Sigma^I_i(\omega)= \Sigma^I_i[F_i(\omega)]$ of the local
  propagator 
  \begin{equation}
    F_i(\omega) = \left[ \omega-\epsilon_i- \sum_{j=1}^K H^{(i)}_j(\omega) \right]^{-1}
  \end{equation}
  which does not contain $\Sigma^I_i(\omega)$.
  With the DMFT self-energy inserted into the equations 
  still the Monte-Carlo scheme applies.
  Since the interaction couples different energies 
  each entry of the sample now represents a local Green's function
  $G_{ii}(\omega)$ on a set of $\omega$ (e.g. Matsubara frequencies).
  Note that during each update of an entry one has to evaluate
  $\Sigma^I_i[F_i(\omega)]$.
  Without disorder, for $\epsilon_i=0$, this scheme reduces to standard
  DMFT.

  The functional dependence 
  $\Sigma^I_i[F_i(\omega)]$ 
  is not explicitly known but requires, despite the approximations made,
  the solution of a quantum-mechanical many-particle problem.
  Solving this model constitutes the main part of a DMFT
  implementation.
  Concerning disordered interacting system the evolved calculation of 
  $\Sigma^I_i[F_i(\omega)]$ poses a severe problem. So far system with
  a finite charge carrier density like a disordered Hubbard model could
  only be addressed in limiting cases~\cite{dobro98}.
  For a single electron in the Anderson-Holstein model we are
  fortunate to explicitly know the functional
  $\Sigma^I_i\left[F_i(\omega)\right]$, given as an infinite continued
  fraction for $T=0$~\cite{Su74,CDF97}:
  \begin{equation}
    \Sigma^I_i(\omega)= \frac{\displaystyle 1 \varepsilon_p \omega_0}
	  {\displaystyle \left[F_i(\omega-1 \omega_0)\right]^{-1} - 
	    \frac{\displaystyle 2 \varepsilon_p
	  \omega_0}{\displaystyle \left[F_i(\omega-2
	  \omega_0)\right]^{-1}  - \frac{\displaystyle 3 \varepsilon_p
	  \omega_0}{\displaystyle \cdots}}}\,.
  \end{equation}
  The $p$-th level of this continued fraction corresponds to the
  emission of $p$ virtual phonons, shifting the energy argument of
  $F_i$ by $p\omega_0$.
  Originally this expression has been obtained as an extension of the CPA
  to describe dynamical interaction effects (concerning dynamical CPA, see~\cite{Su74}).
  In the spirit of the CPA the interaction can be mapped onto an effective
  Hamiltonian which is related to the original problem via a
  consistency condition on the Green's function~\cite{BSB01}.
  This mapping can be combined with the LD construction on the Bethe
  lattice whereby the consequences of the approximations concerning
  interaction and disorder become very explicit~\cite{A03}.
  It turns out that the interaction induced correlations between local
  Green's functions are almost treated on the same level of
  correctness as the disorder induced correlations.
  In particular one finds that the extended LD approach comprises
  cooperative effects which arise from the mutual interplay between
  interaction and disorder.
  This is why the LD approach excels
  approaches which try to replace the interacting disordered system
  by an effective non-interacting disordered system.
  The interaction is there mimicked by effective
  transfer integrals or local potentials,
  but the feedback of disorder on interaction is not accounted for.
  The same objection applies if the interacting disordered system
  is replaced by an effective interacting system, mimicking disorder
  by effective coupling constants.
  The price to be paid on using the LD approach 
  is the rather involved numerical implementation
  combining a repeated solution of the DMFT problem with a Monte-Carlo
  scheme.

  \subsection{Polaron localisation}

  Within the extended scheme just described
  we can study how the electron is affected by disorder and
  EP interaction.
  We will only touch on this question 
  and focus on two significant cases where 
  we compare parts of a phase diagram for polaron localisation to that
  of the Anderson model.
  The reader should be aware that we encounter a very 
  complicated physical situation.
  Already without disorder the Holstein EP interaction 
  displays rich, and very distinct, physics in different parameter regimes.
  With disorder, cooperative effects 
  like the formation of polaron defect states, come into play.
  These effects do hardly fit into a `universal' phase diagram of
  polaron localisation but demand a more concrete description, 
  depending on the specific case studied.
  A more detailed discussion of this and related issues can be found in~\cite{baf04}.
  
\begin{figure}
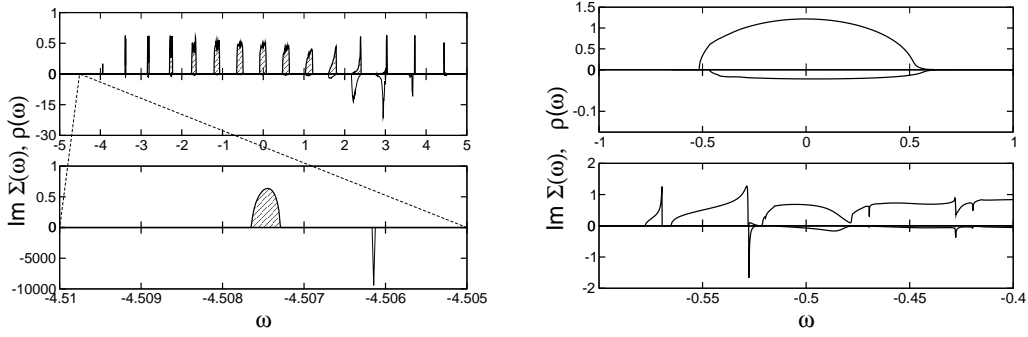

  \hspace*{1cm}
  \includegraphics[width=0.4\linewidth]{HMantiDOS}
  \hspace{0.5cm}
  \includegraphics[width=0.4\linewidth]{HMadiabDOS}
  \caption{DOS $\rho(\omega)$ without disorder ($\gamma=0$),
    obtained within DMFT. $\mbox{Im}\,\Sigma(\omega)$ is displayed
    downwards.  
    Left panel: 
    The strong-coupling, large phonon-frequency case
    $\tilde{\lambda}=9.0,\, \tilde{\omega}_0=0.5625$  can be understood
    from the atomic limit (independent boson model), treating $t$ as a
    perturbation. The DOS shows narrow bands which are separated by 
    $\omega_0$.
    The bandwidth of the lowest band (magnified in the bottom row) is
    $W/W_0 = 3.45 \times 10^{-4}$, i.e. the polaron being extremely heavy.
    Right panel: 
    For weak coupling ($\tilde{\omega}_0=0.05$,
    $\tilde{\lambda}=0.25$, upper row)
    EP interaction essentially leads to inelastic motion of the electron
    ($\mathrm{Im}\,\Sigma(\omega) <0$).
    For stronger coupling ($\tilde{\lambda}=1.0$, same
    $\tilde{\omega}_0=0.05$) a polaron has formed. 
  The lowest polaron band ($W / W_0 = 8.123 \times 10^{-3}$) is
  asymmetric revealing the different character of the polaron to the
  lower and upper band edge. }
   \label{HMDOS}
\end{figure}

\begin{figure}
  \begin{minipage}{0.65\linewidth}
    \hspace*{1cm}
    \includegraphics[width=0.7\linewidth]{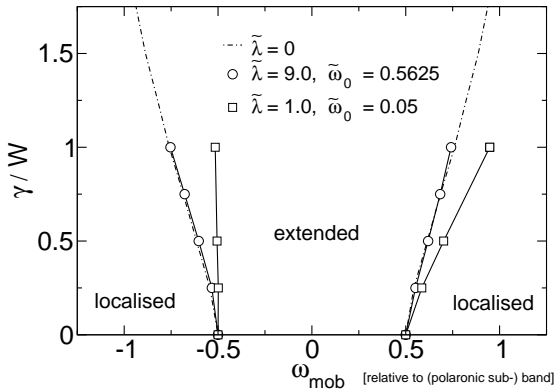} 
  \end{minipage}
  \hfill
  \begin{minipage}{0.34\linewidth}\vspace*{.8cm}
    \caption{In addition to the phase diagram of the Anderson model
      the mobility edges for the two polaron bands from Fig.~\ref{HMDOS} are
    shown.
    The mobility edges are drawn relative to the sub-band position, 
    and disorder is measured in units of the sub-band width $W$~\cite{baf04}.}
    \label{AHM}
  \end{minipage}
\end{figure}

  In Fig.~\ref{HMDOS}
  the DOS for two choices of polaron parameters as obtained by DMFT is
  shown.
  We used the $\eta \to 0$ -- criterion to obtain the mobility edges for
  the lowest polaron sub-band in each of the two cases.
  The comparison to the mobility edges of the Anderson model
  (Fig.~\ref{AHM}) demonstrates that only in special cases 
  the localisation properties of the polaron can be understood 
  as the mere result of the mass renormalisation of the quasiparticle.

  For strong-coupling and large phonon-frequency 
  $\tilde{\lambda}=9.0,\, \tilde{\omega}_0=0.5625$ 
  (circles in Fig.~\ref{AHM}), 
  the polaron
  in the lowest band is basically a heavy particle
  which moves fully coherent but within an extremely narrow band.
  It localises like a bare 
  electron with a rescaled bandwidth, and the internal structure of the
  polaron does not play a role.
  Due to the strong coupling the relevant energy scale for
  localisation has however changed by four orders of magnitude,
  making the polaron extremely susceptible to disorder.

  For intermediate coupling and small phonon-frequency
  $\tilde{\lambda}=1.0$, $\tilde{\omega}_0=0.05$ (squares in Fig.~\ref{AHM}),
  the polaron motion
  is, already without disorder, different at the lower and upper band edge.
  At the lower band edge the polaron is rather mobile,
  while at the upper band edge it tends to be immobile.
  Concerning the localisation properties the polaron at the lower band
  edge is thus almost unaffected by disorder and does not readily localise.
  At the upper band edge, in contrast, it is easily localised.
  Note that the lowest polaron band is again coherent: localisation is
  not just affected by incoherent motion but intricately depends on
  the internal structure of the polaron. 
  Be also aware that disorder is measured in units of the
  renormalised band width: the polaron at the lower band edge is more
  difficult to localise than a bare electron only with respect to the
  relevant energy scale $W$, i.e. the polaronic bandwidth.

  In Fig.~\ref{AHM} we have only shown data for moderate
  disorder, i.e. below $\gamma_c(\omega=0)$ when, for the bare electron,
  all states become localised.
  In the first (strong-coupling) case we can 
  indeed localise all states in the lowest polaron band 
  and obtain the respective $\gamma_c(\omega=0)$ which has the
  value of the bare electron renormalised by $W/W_0$.
  Here disorder induced mixing with excited polaron states 
  is prevented by the large gap to the next band ($W \ll \omega_0$).
  In the second (adiabatic intermediate-coupling) case, however, with increasing disorder the lowest
  band merges with the next polaron band before it is completely
  localised.
  Then the relevant energy scale for $\gamma$ changes by one order of
  magnitude to the joint bandwidths of the two lowest bands.
  Hence an equivalent to $\gamma_c(\omega=0)$ does not exist.
  Apparently we can only draw parts of a phase diagram for
  polaron localisation. Again, EP interaction changes the
  localisation properties in a more complicated way than
  thinking only in terms of mass renormalisation would suggest.
 
\section{Summary}    

In disordered systems Green's functions emerge as
stochastic, random quantities.
It is reasonable to take this stochastic character serious
and to incorporate distributions of those quantities 
in a description of disorder.
Following~\cite{Abou73} the proposed LD approach is based 
on distributions of the local Green's function.
Being exact on a Bethe lattice, approximate on general lattices,
it accounts for quantum interference leading to Anderson localisation,
and multiple scattering on clusters like in a binary alloy.

The LD approach is an extension of the CPA including fluctuations.
Conversely, it contains the CPA in the limiting case when fluctuations
are replaced by averages. We can then understand that
the CPA results are good if  -- and only if -- the distribution
of the LDOS is characterised by its average.
This condition does not mean that the disorder is weak:
Even for small $\gamma$ in the Anderson model states at the band edges
are localised, and beyond the scope of the CPA.
A similar restriction exists for the minority band in the binary alloy
model.

The great challenge is however not disordered, 
but interacting disordered systems.
Here the LD approach can provide a description of the microscopic
interplay between interaction and impurity scattering.
Locally, both interaction and impurity scattering are treated in the
best approximation possible:
DMFT is the best single-site approximation to interaction;
and impurity scattering is exactly treated (on general lattices at
least locally exact).

The LD approach depends on the accuracy of the DMFT self-energy.
A deep understanding of this part of the problem is important.
How the involved numerical methods (quantum Monte-Carlo, exact diagonalisation,
density matrix renormalisation group, to mention a few)
or approximative schemes (e.g. iterative perturbation theory (IPT)) 
used for the
calculation of the DMFT self-energy~\cite{georges96}
cope with the considerable fluctuations in the local propagator $F_i(\omega)$
induced by disorder is by no means clear
(e.g. the perturbative expansion in the IPT is best for weakly varying
$F_i(\omega)$).
Since we largely understand the LD approach's application to disorder,
how to guarantee for the quality of these methods is still the biggest
problem  -- the computational costs for their implementation in the
Monte-Carlo scheme notwithstanding.
Our results for the localisation of a single Holstein polaron, where
these objections are absent, demonstrate that the LD approach by
itself allows for a comprehensive description and detailed study 
of interacting disordered systems.
Physically we learned how intricate and rich the interplay of
interaction and localisation can be.
Future work will certainly support and extend this observation.

\ackn
We would like to thank F.~X.~Bronold for valuable discussions.
\vfill
  
\section*{References}

\providecommand{\newblock}{}

\end{document}